\documentclass[twocolumn,aps,prb]{revtex4-1}
\usepackage[latin9]{inputenc}
\usepackage{color}
\usepackage{verbatim}
\usepackage{bm}
\usepackage{amsmath}
\usepackage{graphicx}
\usepackage{bm,amssymb}
\usepackage{hyperref}
\usepackage{placeins}
\hypersetup{
    colorlinks=true,       
    linkcolor=teal,         
    citecolor=blue,        
    filecolor=magenta,  
    urlcolor=cyan          
}
\usepackage[dvipsnames]{xcolor}

\makeatletter

\providecommand{\tabularnewline}{\\}

\@ifundefined{textcolor}{}
{%
 \definecolor{BLACK}{gray}{0}
 \definecolor{WHITE}{gray}{1}
 \definecolor{RED}{rgb}{1,0,0}
 \definecolor{GREEN}{rgb}{0,1,0}
 \definecolor{BLUE}{rgb}{0,0,1}
 \definecolor{CYAN}{cmyk}{1,0,0,0}
 \definecolor{MAGENTA}{cmyk}{0,1,0,0}
 \definecolor{YELLOW}{cmyk}{0,0,1,0}
}

\usepackage{bm}
\usepackage{color}
\usepackage{placeins}
\usepackage{subfigure}
\usepackage{comment}
\usepackage{braket}

\def\be{\begin{equation}}       \def\ee{\end{equation}}
\def\bea{\begin{eqnarray}}      \def\eea{\end{eqnarray}}
\def\ba{\begin{array} }
\def\ea{\end{array} }
\def\bnum{\begin{enumerate} }
\def\enum{\end{enumerate}}

\def\=>{\Rightarrow}
\def\>{\rightarrow}

\def\eye2{Fathbb{I}}

\def\k{\vec{k}}

\def\d0{\Delta_{0}}

\def\ea{\ensuremath \varepsilon_{B_{1g}}}

\def\k{\ensuremath \bm{k}}
\def\kp{\ensuremath \bm{k}^{\prime}}

\newcommand{\nocontentsline}[3]{}
\newcommand{\tocless}[2]{\bgroup\let\addcontentsline=\nocontentsline#1{#2}\egroup}

\makeatother

\begin{document}

\title{ Transverse fields to tune an Ising-nematic quantum
critical transition }

\author{Akash V. Maharaj$^{1,2}$, Elliott W. Rosenberg$^{3}$, Alexander
T. Hristov$^{1,2}$, Erez Berg$^{4,5}$, Rafael M. Fernandes$^{6}$,
Ian R. Fisher$^{2,3}$, and Steven A. Kivelson$^{1,2}$}

\affiliation{${}^{1}$Department of Physics, Stanford University, Stanford, CA 94305,
USA}

\affiliation{${}^{2}$Stanford Institute for Materials and Energy Sciences, SLAC National Accelerator Laboratory, Menlo Park, California 94025, USA}

\affiliation{${}^{3}$Department of Applied Physics, Stanford University, Stanford,
CA 94305, USA}

\affiliation{${}^{4}$Department of Condensed Matter Physics, Weizmann Institute of Science, Rehovot, Israel 761001}

\affiliation{${}^{5}$Department of Physics, University of Chicago, Chicago, IL 60637, USA}

\affiliation{${}^{6}$School of Physics and Astronomy, University of Minnesota, Minneapolis,
MN 55455, USA}

\date{\today }

\maketitle
\textbf{
The paradigmatic example of a continuous quantum phase transition
is the transverse field Ising ferromagnet. In contrast to classical
critical systems, whose properties depend only on symmetry and
the dimension of space, the nature of a quantum phase transition
also depends on the dynamics\cite{sachdev2007quantum,sondhiRMP97}. In the transverse field Ising model,
the order parameter is not conserved and increasing the transverse 
field enhances quantum fluctuations until they become strong enough 
to restore the symmetry of the ground state. Ising pseudo-spins can 
represent the order parameter of any system with a two-fold 
degenerate broken-symmetry phase, including electronic nematic 
order associated with spontaneous point-group symmetry breaking\cite{kivelsonNature98}. 
Here, we show for the representative example of orbital-nematic 
ordering of a non-Kramers doublet that an orthogonal strain or a 
perpendicular magnetic field plays the role of the transverse
field, thereby providing a practical route for tuning appropriate
materials to a quantum critical point. While the transverse fields are 
conjugate to seemingly unrelated order parameters, their non-trivial 
commutation relations with the nematic order parameter, which can 
be represented by a Berry-phase term in an effective field theory\cite{SenthilPRB06}, 
intrinsically intertwines the different order parameters\cite{fradkinRMP15}.}

Ising nematic states are increasingly recognized as important
members of the phase diagrams of several families of strongly correlated
electronic materials. In the iron-based superconductors,
electronic nematic phases are unambiguously present\cite{chuangScience2010,chuScience2010,duszaEPS2011,yiPNAS2011,luScience2014,fernandesNatPhys2014}, and the maximum
superconducting transition temperature often correlates with the location
of a putative quantum critical point with a nematic character. In
the cuprate superconductors, there is a wealth of indirect evidence\cite{andoPRL2002,howaldPRB2003,hinkovScience2008,lawler2010intra,cyr2015two}
that electronic anisotropy is enhanced in the underdoped regions of
the phase diagram, with a possible quantum critical point near optimal
doping. 
As it is breaks a discrete rotational symmetry, Ising-nematic 
order is relatively robust to disorder and low dimensionality.
Moreover, nematic critical fluctuations have zero momentum, and so couple to (nearly) all low energy fermionic quasiparticles.
Consequently, there has been considerable theoretical interest 
in the role of nematic quantum critical fluctuations as a route to non-Fermi liquid metallic behavior and even superconductivity\cite{kim2004pairing,Maier2014pairing,lederer, metlitski2015Cooper, fernandesPRL2016}.

\begin{figure}
\centering \includegraphics[width=0.48\textwidth]{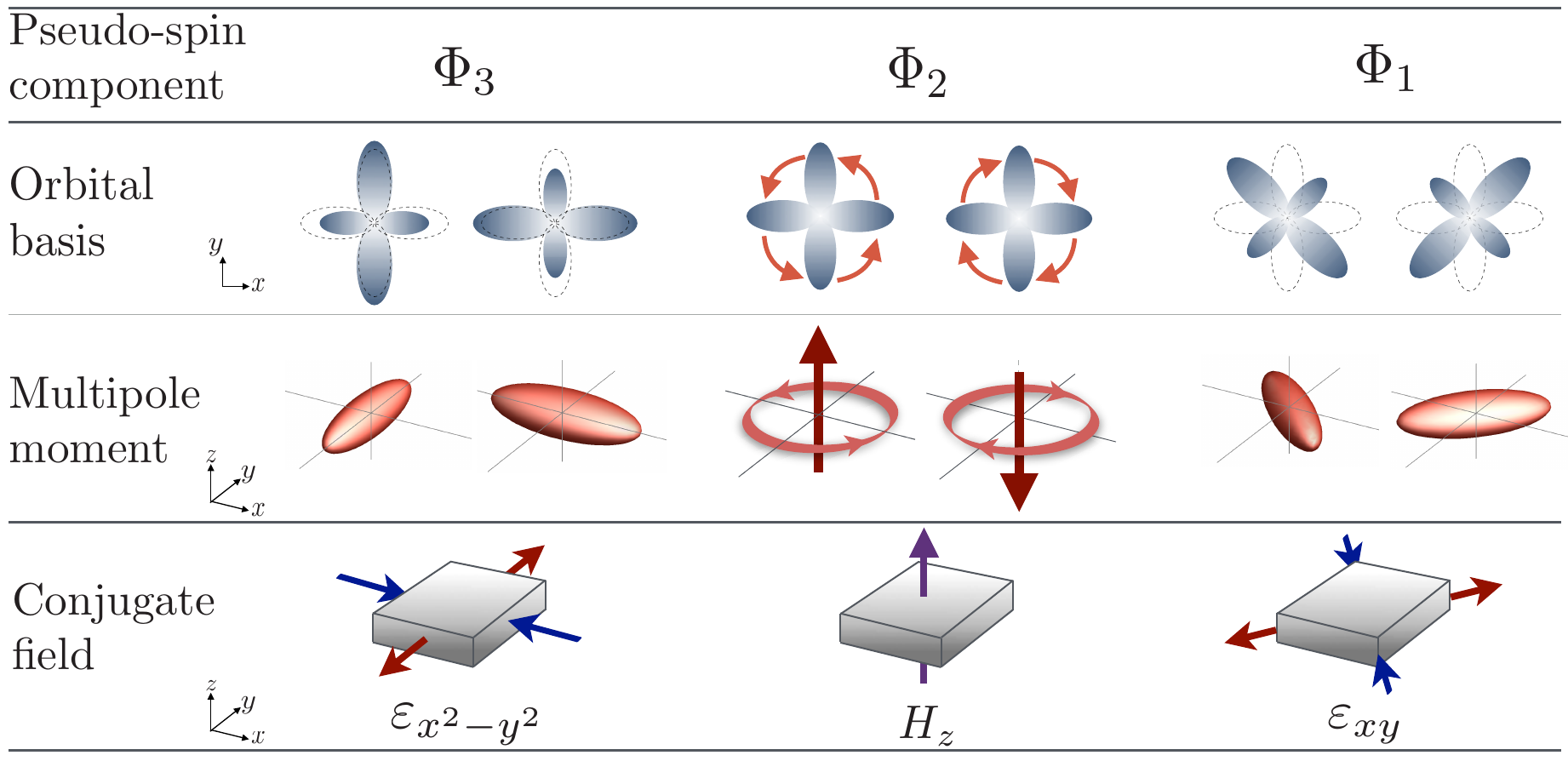}
\caption{\textbf{Nematic order as one component of pseudo-spin.} Schematic representation of the three types of order possible for
a non-Kramers $E_{g}$ doublet in a tetragonal system, corresponding
to three components of a pseudo-spin representation. Specifically,
we consider ordered states at wavevector $\mathbf{Q}=0$, \textit{i.e.}
ferro-orbital order. The second row illustrates the polarization of
local (Wannier) orbitals for a model comprising orbitals with $xz$
and $yz$ symmetry. Pairs of states represent opposite signs/
polarization of the Ising pseudospins The longitudinal component
$\Phi_{3}$ corresponds to orbital polarization with an $x^{2}-y^{2}$
($B_{1g}$) symmetry; $\Phi_{1}$ corresponds to an orbital polarization
with an $xy$ ($B_{2g}$) symmetry; and $\Phi_{2}$ corresponds to
an orbital magnetic moment. $\Phi_{3}$ and $\Phi_{1}$ are clearly
realizations of Ising nematic order. The corresponding multipole moments
and associated conjugate fields are shown in the third and fourth
rows for each pseudo-spin component.}
\label{fig:tablea} 
\end{figure}

Identifying appropriate means to tune systems through a continuous
nematic quantum phase transition is therefore of considerable importance.
Previous studies have tuned nematicity through magnetic
fields, hydrostatic pressure, or chemical composition. The latter
two parameters, however, cannot be varied continuously
during an experiment. Moreover, chemical substitution frequently
changes the band filling and inevitably affects the degree of disorder,
potentially confounding attempts to delineate the roles played by
individual variables. Here, we demonstrate that orthogonal strain
or perpendicular magnetic fields promote quantum fluctuations that
suppress a nematic phase transition, opening a promising
avenue to tune to a nematic quantum critical point.


This effect bears a close resemblance to the physics
of the transverse-field Ising model. To construct the nematic pseudo-spins,
consider a tetragonal material with electronic nematic
order associated with the splitting of a non-Kramers doublet.
The corresponding Wannier functions of the orbital doublet at each
site transform according to the $E_{g}$ irreducible representation,
and therefore have either $xz$ symmetry (represented by the orbital
index $a=x$) or $yz$ symmetry (with orbital index $a=y$). To simplify
the discussion, and without loss of generality, we consider spinless electrons. In the Hilbert space
corresponding to a single site $\vec{R}$ with an orbital doublet,
any operator can be expressed as a linear combination of the total
number operator and the vector pseudo-spin operators 
\begin{equation}
\Phi_{\alpha}
(\vec{R})\equiv\sum_{a,a^{\prime}}c_{a,\vec{R}}^{\dagger}\tau_{a,a^{\prime}}^{(\alpha)}c_{a^{\prime},\vec{R}}^{\vphantom{\dagger}}\:,\label{pseudospin}
\end{equation}
where $c_{a,\vec{R}}^{\dagger}$ creates an electron in the 
Wannier orbital of symmetry $a$, $\tau_{a,a^{\prime}}^{(\alpha)}$
are the Pauli matrices, and $\alpha=1,$ 2, 3. 
The components $\Phi_{\alpha}(\vec{R})$  
satisfy the same commutation relations as a spin operator, 
\begin{equation}
\left[\Phi_{\alpha}(\vec{R}),\Phi_{\beta}(\vec{R}^{\prime})\right]=
i\ \delta_{\vec{R},\vec{R}^{\prime}}\ \epsilon_{\alpha\beta\gamma}\ \Phi_{\gamma}^ {}(\vec{R})\label{commutation}
\end{equation}
where $\epsilon_{\alpha\beta\gamma}$ is the Levi-Civita symbol. In
terms of the orbital density $\hat{n}_{a}\equiv c_{a,\vec{R}}^{\dagger}c_{a,\vec{R}}$,
$\Phi_{3}$ can be identified as the difference in occupancy of $x$ and $y$ orbitals
(Figure 1)
\begin{eqnarray}
\Phi_{3}(\vec{R})\equiv\hat{n}_{x}(\vec{R})-\hat{n}_{y}(\vec{R}).
\end{eqnarray}

We identify $\Phi_{3}$ as the nematic order parameter that breaks the equivalence
between the $x$ and $y$ axis; in group-theory language,  
it has $B_{1g}$ ($x^{2}-y^{2}$) symmetry. The phase in which
$\left\langle \Phi_{3}\right\rangle \neq0$ breaks tetragonal symmetry
but preserves horizontal and vertical mirror symmetries. 
By changing the orbital basis, it is easy to see that $\Phi_{1}$
is associated with breaking the equivalence between
the two diagonals $x\pm y$, and corresponds to a nematic order parameter with $B_{2g}$ ($xy$)
symmetry, which breaks tetragonal symmetry but preserves diagonal
mirror symmetries. Finally, $\Phi_{2}$ is identified with the difference
in occupancy of $x\pm iy$ orbitals, and corresponds
to an orbital magnetic moment that preserves tetragonal symmetry but
breaks time-reversal symmetry
(see Figure \ref{fig:tablea}).

Expressed in terms of local multipole moments, $\Phi_{3}$ and $\Phi_{1}$
correspond to electric quadrupole moments, while $\Phi_{2}$ corresponds
to a magnetic dipole moment, immediately delineating the appropriate
conjugate fields for each of the three components of the vector operator,
as shown in Figure~\ref{fig:tablea}. The combination of time-reversal and
point group symmetries forbid terms in the Hamiltonian,
$\mathcal{H}$, that are odd functions of $\Phi_{\alpha}$.
However, application of symmetry breaking strains $\varepsilon_{xx}-\varepsilon_{yy}$
(corresponding to unequal lattice distortions along $x$ and $y$)
and $\varepsilon_{xy}$ (corresponding to a shear distortion of the
lattice) or a uniform magnetic field perpendicular to the plane $H_{z}$
induce linear terms of the form 
\begin{equation}
\mathcal{H}\to\mathcal{H}-\sum_{\vec{R}}{\bm{h}}\cdot{\bm{\Phi}}(\vec{R})\label{coupling}
\end{equation}
where, to leading order: 
\begin{eqnarray}
h_{3} &  & =\lambda_{3}\left(\varepsilon_{xx}-\varepsilon_{yy}\right)+\ldots\nonumber \\
h_{1} &  & =\lambda_{1}\varepsilon_{xy}+\ldots \\
\ h_{2} &  & =\lambda_{2}H_{z}+\ldots\nonumber\label{fields}
\end{eqnarray}

\begin{figure}
\centering \includegraphics[width=0.48\textwidth]{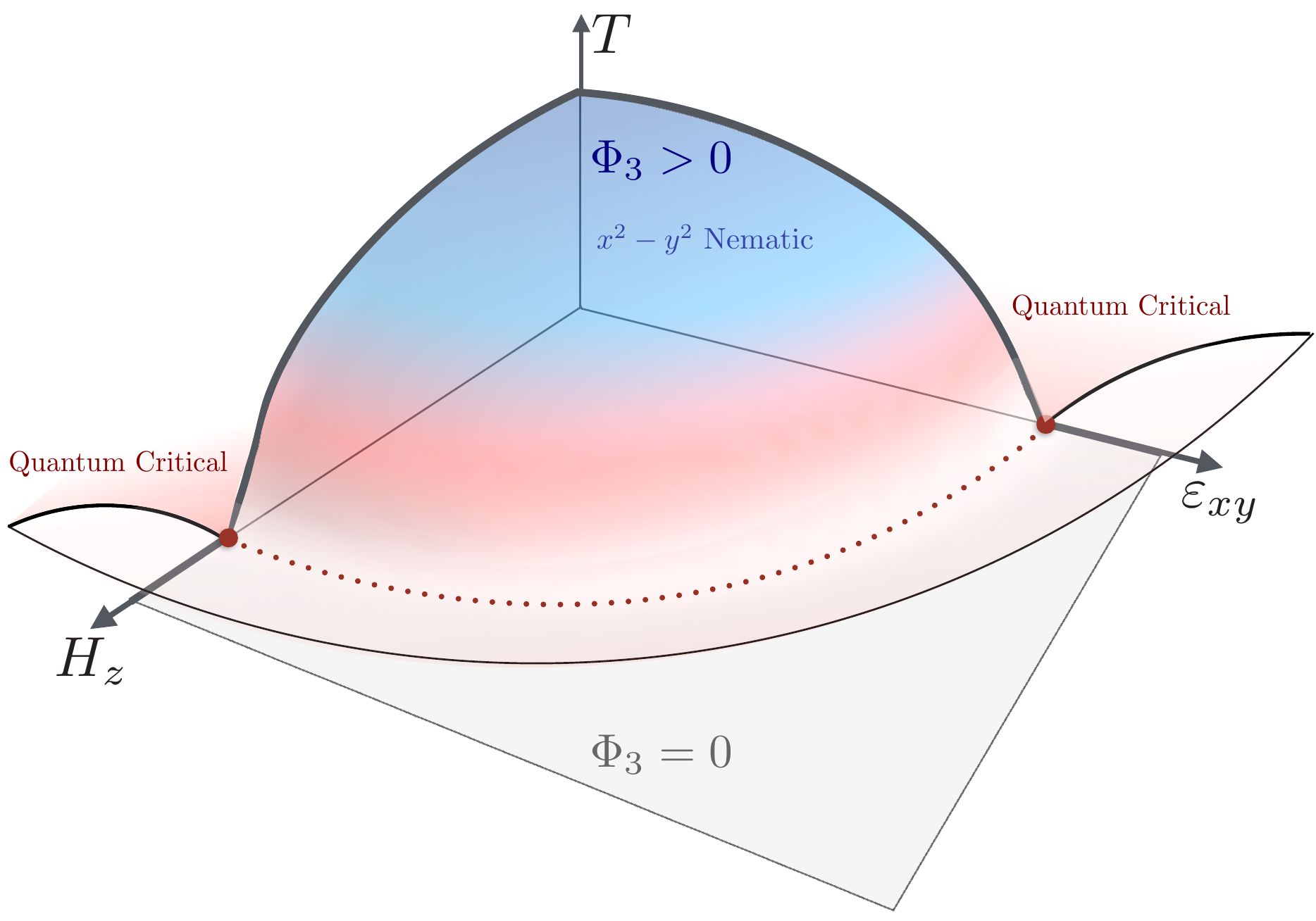}
\caption{\textbf{Nematic quantum criticality induced by transverse fields.} 
Here, for $x^2-y^2$ nematic order, the transverse fields are shear strain $\varepsilon_{xy}$ and a perpendicular magnetic
field, $H_{z}$; increasing their magnitudes strengthens quantum fluctuations,
which drive the finite temperature nematic phase transition to a quantum critical 
line in the $H_{z}-\varepsilon_{xy}$ plane
at $T=0$ (dotted red line). The quantum critical fan is extended to a ring of critical
behavior (shaded in red). 
}
\label{fig:cartoon1} 
\end{figure}

The analogy with the transverse field Ising model is now
apparent. The ``longitudinal'' field, $h_{3}$,
is a symmetry breaking field: in a phase 
with $x^{2}-y^{2}$ symmetry, even an infinitesimal
$h_3$  lifts the two-fold degeneracy of the state,
selecting the phase in which $h_{3}\langle\Phi_{3}\rangle>0$,
so a finite $h_{3}$ necessarily smears the nematic
transition. On the other hand, the other two components
of ${\bm{h}}$ behave as ``transverse'' fields. They reduce the
symmetry of the Hamiltonian while preserving the symmetry $\Phi_{3}\to-\Phi_{3}$, which permits a well-defined nematic
transition in the presence of non-zero $h_{1}$ and/or $h_{2}$.
However, large enough values of the transverse fields
will preclude nematic order, since the commutation relations
in Eq. \ref{commutation} imply that a pseudo-spin with a well-defined
value of $\Phi_{1}$ or $\Phi_{2}$ must be highly uncertain in $\Phi_{3}$.
This makes it possible to induce a $B_{1g}$ nematic quantum
phase transition by applying shear strain or magnetic field (see Fig.
\ref{fig:cartoon1}).

 \label{sec:4f}

An informative realization of Ising nematic order, which clearly illustrates the effect of transverse fields, is ferroquadrupolar order in $4f$ intermetallic compounds\cite{GehringGehring, Morin1990Quadrupolar, Santini4fRevModPhys}.
In their trivalent state, the rare earth elements Ce - Yb 
have a partially filled $4f$ orbital whose small extent 
implies that its electronic states are effectively localized.  
A hierarchy of energy scales then
determines the character of the ground state. Focusing first on a
single site, Hund's rules in the limit of strong spin-orbit
coupling dictate that the total angular momentum $J$ is
the only good quantum number. The local crystal electric field (CEF) then acts as a perturbation,
splitting the $(2J+1)$-fold degenerate Hund's rule ground state.
The CEF eigenstates are linear combinations of these states 
with their character determined by the point
group symmetry\cite{GehringGehring}. For the specific case of intermetallic systems the additional conduction
electrons mediate not only an effective interaction between the local
magnetic moments but also between the local charge distributions via
a generalization of the Ruderman-Kittel-Kasuya-Yosida (RKKY) exchange mechanism\cite{Schmitt1985,MorinSchmitt88,SchmittQuadrupolar90, Morin1990Quadrupolar}. At the
two ``ends'' of the $4f$ series (corresponding to Ce
and Pr on the left, or Tm and Yb on the right), Hund's rules necessarily
imply that these elements have relatively large orbital angular momenta
(and hence large electric quadrupole moments), but relatively
small total spin (and hence small magnetic exchange energies). For
these elements, quadrupole-quadrupole interactions can 
exceed spin-spin interactions. Materials with such constituents can  
exhibit ordered phases in which the local $4f$ orbitals develop a
spontaneous quadrupole moment at a higher temperature than any long
range magnetic order. 

\begin{figure}
\centering \includegraphics[width=0.45\textwidth]{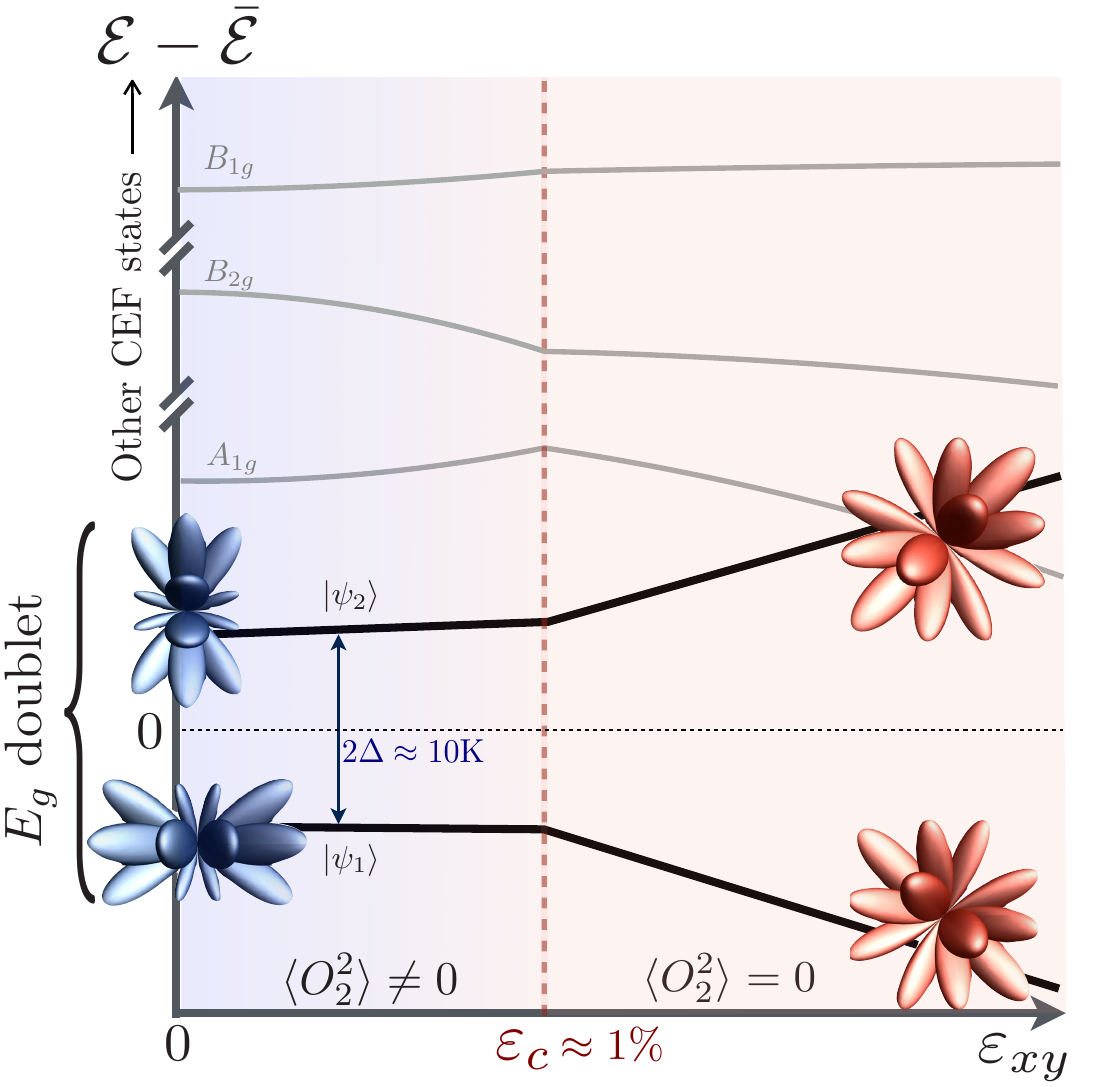}
\caption{\textbf{Strain-induced nematic quantum phase transition in TmAg$_2$.} A mean field calculation of the effect of shear strain $\varepsilon_{xy}$ on the energy eigenvalues $\mathcal{E}$ and charge distribution of the lowest CEF eigenstates of the $4f$ intermetallic compound TmAg$_2$ which has undergone ferroquadrupole (nematic) order with a $B_{1g}$ symmetry. The two lowest energy eigenvalues correspond to an $E_g$ doublet that has been split due to quadrupole-quadrupole interactions. As discussed in the main text, this $E_g$ doublet may be formally identified with a pseudospin-1/2, and so shear strain $\varepsilon_{xy}$ acts as a transverse field and reduces the spontaneous quadrupole moment $\langle \hat{O}^2_2 \rangle$ by inducing quantum fluctuations between the eigenstates $\ket{\psi_1}$ and $\ket{\psi_2}$. At a critical strain $\varepsilon_c \approx 1\%$, the nematic order is completely suppressed at a quantum phase transition. The color scale represents the mixing angle between $\ket{\psi_1}$ and $\ket{\psi_2}$. See Supplemental Material for a full description of the mean-field model used to calculate the strain dependence. Note that for clarity, we have subtracted the average energy of the $E_g$ doublet, $\bar{\mathcal{E}}$ from each energy eigenvalue.}
\label{fig:energylevels} 
\end{figure}

This is precisely the case for the tetragonal intermetallic
compounds TmAg$_{2}$ and TmAu$_{2}$, which display ferroquadrupolar
order with $B_{1g}$ symmetry, below a critical temperature of approximately 5K and 7K respectively\cite{MorinTMAg2PRB93,KosakaTmAu2PRB98, MorinTmAu2JOP99}. 
Near the ferroquadrupolar transition, the only states which are both thermally populated and significant to the transition
are the two CEF eigenstates belonging to the non-Kramers doublet, labeled ``$E_{g}$ doublet'' in Fig.~\ref{fig:energylevels}. 
The states of this doublet can be treated as a pseudo-spin $1/2$; projecting the Stevens operators $O_{2}^{2}=J_{x}^{2}-J_{y}^{2}$,
$P_{xy}=(J_{x}J_{y}+J_{y}J_{x})$, and $J_{z}$ to this doublet yields operators with the same commutation relations
defined in Eq.~\ref{commutation} (see supplementary material for more details). 
Consequently, $\varepsilon_{xy}$ and $H_z$ are appropriate 
transverse fields to tune the ferroquadrupolar transition
to a quantum critical point. Indeed, a magnetic field oriented
along the crystalline $c$-axis has been shown to suppress quadrupole
order\cite{MorinTMAg2PRB93,MorinTmAu2JOP99}, though the effect of shear strain predicted here has yet to be demonstrated.

Our preceding discussion focused on nematic systems whose 
relevant low-energy states are a non-Kramers doublet. 
In this case, we emphasized the commutation relations 
in Eq. \ref{commutation}. These imply the transverse field enhances quantum
fluctuations of the order parameter and, when sufficiently large, drives 
the nematic transition temperature $T_{c}\to0$.
However, this result does not generalize to all possible types of Ising 
nematic order, as symmetry does not guarantee the availability of an external, experimentally 
applicable transverse field.

Consider, for example, the case of a tetragonal, single band
metal undergoing a Pomeranchuk transition at which the Fermi surface 
undergoes a symmetry-breaking nematic distortion. In this case, an 
$x^{2}-y^{2}$ nematic order parameter must involve at least nearest-neighbor sites, 
\begin{equation}
\Phi_{3}(\vec{R})\equiv\sum_{\vec{R}^{\prime}}f(\vec{R}-\vec{R}^{\prime})\left[c_{\sigma,\vec{R}}^{\dagger}c_{\sigma,\vec{R}^{\prime}}+{\rm H.C.}\right]\label{eq:1bandDwave}
\end{equation}
where $f$ is the ``$d$-wave'' form factor: $f(\pm\hat{x})=-f(\pm\hat{y})=1$
and $f(\vec{R})=0$ otherwise. From a symmetry perspective, this 
is no different than the situation analyzed above. Indeed,  
a field with the symmetry of $\Phi_{1}$  
can be constructed by choosing $f(\vec{R}-\vec{R}^{\prime})$
to have $xy$ character in Eq. \ref{eq:1bandDwave}.

However, because the system has only a single $s$-orbital, it is
not possible to write down an orbital magnetic moment, i.e. there
is no $\Phi_{2}
(\vec{R})$. More importantly, while the commutation relations of the nematic
operators are moderately complicated
, they commute when averaged over sites. 
\begin{equation}
\left[
\Phi_{3}(\vec{R}),\sum_{\vec{R}^{\prime}}\Phi_{1}(\vec{R}^{\prime})\right]=0.\label{commute}
\end{equation}

Consequently, although external strain still couples
to $\vec{\Phi}$ as in Eqs. \ref{coupling} and \ref{fields} (with
$h_{2}=0$), Eq. \ref{commute} implies that $h_{1}$ does not enhance
quantum fluctuations of $\Phi_{3}$. Of course, changing the value
of any term in the microscopic Hamiltonian which does not explicitly
break a relevant symmetry will generally result in a shift in $T_{c}$.
However,  while $T_{c}$ will be a parametric function of $h_{1}$, it
does not necessarily decrease for increasing strain.

To illustrate this crucial point, we calculated
the Hartree-Fock phase diagram of such a one-band model undergoing
a Pomeranchuk transition characterized by a non-zero $\Phi_{3}$ (see
Supplementary Material). Application of shear strain (i.e.
a finite $h_{1}$) decreases the second-order transition temperature. However,
as it approaches $T=0$, the nematic transition becomes generically
first-order, preempting a quantum critical point.
By contrast, when the system undergoing the Pomeranchuk transition
has two bands arising from $x$ and $y$ orbitals, the Hartree-Fock
phase diagram gives a nematic transition that remains second-order
all the way to $T=0$.

As a first step toward a more general formulation
of the problem, which highlights the role of purely quantum effects,
it is instructive to consider the effective field theory of the pseudo-spin
$\vec{\Phi}$. We focus on an insulating tetragonal system with an
odd number of sites. If there is a (non-Kramers) orbital doublet that
transforms according to a two-dimensional representation ($E_{g}$
or $E_{u}$) of the tetragonal point group, then there is a local
pseudo-spin-1/2 associated with each unit cell of the crystal. In
this case, all the states \textendash{} including the ground-state \textendash{} are at least
doubly degenerate. It is illuminating to draw an analogy with
the insulating spin-1/2 \emph{antiferromagnet}:
in that case, the two-fold degeneracy of every state is reflected
in a topological term in the effective field theory. Correspondingly,
the effective field theory for $\vec{\Phi}$ also contains a quantized
Berry-phase term:
\begin{align}
S_{\mathrm{top}}[{\bm{\Phi}}]=\int_{0}^{1/T}d\tau{\bm{A}}({\bm{\Phi}})\cdot\dot{\bm{\Phi}}.
\end{align}
where $\tau$ is the imaginary time and ${\bm{A}}({\bm{\Phi}})$ is
the Berry connection associated with a monopole in ``nematic space''
of topological charge $q$. The two-fold degeneracy, absent for the
case of singlet orbitals, is encoded in the parity of $q$, which
has important consequences for the system. 
While in the insulating antiferromagnet this term distinguishes integer from 
half integer spins, in our ferroquadrupolar (nematic) system it encodes the 
non-trivial commutation relations between $\Phi_1$, $\Phi_2$, and $\Phi_3$.    
Given that $\Phi_1$ and $\Phi_2$ break entirely different symmetries than 
$\Phi_3$, in a usual Landau-Ginzburg-Wilson treatment of the problem, explicit 
reference to these other forms of order would be supressed except near a 
fine-tuned multi-critical point.  In the present problem, these different orders 
are implicitly intertwined by the commutation relations in Eq. 2, such that 
anything that increases one component of $\Phi$ increases the amplitude 
of the quantum fluctuations of the others.


In summary, our work introduces a powerful new tuning
parameter, namely shear strain, that enhances quantum nematic fluctuations
and which can be varied continuously, without introducing disorder,
and without breaking time reversal symmetry. Longitudinal strain has
already been extensively used to probe the nematic susceptibility
in various systems\cite{chuScience2010}.
Transverse strain provides a new way to tune through the nematic 
phase diagram, and especially to access  the quantum critical regime.

For many materials of current interest, the magnitude of the coupling
between nematic and elastic degrees of freedom is not known quantitatively.
However, for the specific examples of TmAu$_{2}$ and TmAg$_{2}$ 
discussed here, estimates of the magnetoelastic coupling coefficient
combined with the measured elastic modulus $c_{66}$ indicate that
shear strains of order 1\% would be sufficient to access
the quantum phase transition (see Sec. IV D of Supplementary Material).
Such strains are certainly not inconceivable, and they might
be smaller in other materials depending on microscopic details.

A metal near a time-reversal symmetric Ising-nematic quantum
critical point is expected to have an enhanced instability towards
superconductivity\cite{kim2004pairing,Maier2014pairing,lederer, metlitski2015Cooper, fernandesPRL2016}. Thus, tuning
the ferroquadrupolar transition of TmAu$_{2}$ and TmAg$_{2}$ to
a quantum critical point using shear strain may drive these compounds
into a superconducting phase. In contrast, a $c$-axis magnetic field
naturally suppresses superconductivity.
%
%
Finally, shear strain 
does not only couple to 
$\mathbf{Q}=0$ ferroquadrupolar order; 
$\varepsilon_{xy}$ may also act as a transverse field for 
antiferroquadrupolar order 
by promoting quantum fluctuations between components of each local pseudo-spin. 


We acknowledge useful conversations with Daniel Agterberg and Andrey Chubukov. 
AVM, ATH, IRF, and SAK were supported by the Department of
Energy, Office of Basic Energy Sciences under contract DE-AC02-76SF00515.
EWR was supported by the Gordon and Betty Moore Foundation EPiQS Initiative
through grant GBMF4414. 
RMF was supported by the U.S. Department of Energy, Office of Science, Basic Energy
Sciences, under Award number DE-SC0012336.
EB was supported by
the Israel Science Foundation under Grant No. 1291/12 and
by the US-Israel BSF under Grant No. 2014209.


\tocless\bibliography{transverse_refs.bib}

\widetext
\clearpage
\begin{center}
\textbf{\large Supplemental Material: Transverse fields to tune an Ising-nematic quantum
critical transition}
\end{center}
\setcounter{equation}{0}
\setcounter{figure}{0}
\setcounter{table}{0}

\makeatletter
\renewcommand{\theequation}{S\arabic{equation}}
\renewcommand{\thetable}{S\arabic{table}}
\renewcommand{\thefigure}{S\arabic{figure}}
\renewcommand{\bibnumfmt}[1]{[S#1]}
\renewcommand{\citenumfont}[1]{S#1}

\makeatletter
\def\l@subsubsection#1#2{}
\makeatother

\tableofcontents

-----------------------------------------------------------------------------------------------------
\FloatBarrier
\section{Role of Berry phase terms in field theoretic formulation}

To achieve a more general understanding of the quantum aspect of the problem, it is useful to consider the problem from the more abstract -- less microscopic -- perspective of an effective field theory.  As a first step in that direction, we can integrate out the microscopic degrees of freedom, leaving us with an effective action, $S[{\bm \Phi}]$,  which determines the quantum statistical mechanics of a three-component collective field, ${\bm \Phi}$, such that $\Phi_3$ is the nematic order parameter of interest. For simplicity, we will assume we are dealing with in insulator, such that $S[{\bm \Phi}]$ can be sufficiently well approximated by a local expression that a conventional Landau-Ginzburg-Wilson (LGW) approach to the problem is reasonable. In contrast, in a metallic system, $S$ might be a complicated non-local functional; a full analysis of this case is beyond the scope of the present work. We will comment on the implications for the metallic case at the end.

For a classical problem, we would simply take  $S=S_{LGW}$ where the usual symmetry analysis would be applied to determine the allowed terms in powers of the fields and their derivatives.  Indeed, if we are not near a fine-tuned multicritical point, we would simply drop (or more formally, integrate out) the remaining components of ${\bm \Phi}$; certainly, if there were no sign of orbital ferromagnetism, we would drop $\Phi_2$. 

But for the quantum problem, there is another possible Berry-phase term that is allowed:
\begin{equation}
S[{\bm  \Phi}] = i\sum_{\vec R}\Theta[{\bm \Phi}_{\vec R}] + S_{LGW}[{\bm \Phi}]
\end{equation}
where
\begin{equation}
\Theta[{\bm \Phi}] = \int_0^\beta d\tau {\bm  A}({\bm \Phi})\cdot \dot{\bm \Phi}.
\end{equation}
Here, ${\bm A}({\bm \Phi})$ is the Berry connection. 

The form of ${\bm A}({\bm \Phi})$ is constrained by symmetry: $A_{1,3}$ must be odd under time reversal, while $A_2$ is even. Under the point group symmetry transformation, the components of $A_{1,2,3}$ transform in the same way as $\Phi_{1,2,3}$, respectively. In addition, ${\bm A}({\bm \Phi})$ can have singularities at a discrete set of points. Namely, the Berry curvature $\mathcal{B} \equiv {\bm \nabla}_{\mathbf{\Phi}} \times  {\bm A}$ satisfies
\begin{equation}
{\bm \nabla}_{\mathbf{\Phi}} \cdot \mathcal{B} = \sum_i \frac{q_i}{2} \delta({\bm \Phi} - {\bm \Phi}_i),
\end{equation}
where ${\bm \Phi_i}$ is a discrete set of points in ${\bm \Phi}$ space and $q_i$ are a set of integers. (The quantization of $q_i$ comes from the requirement that the amplitude of the path integral is single valued.) 

A ``monopole'' of ${\mathcal{B}}$ at ${\bm \Phi}_i$ must be accompanied by symmetry-related partners under mirror reflections relative to the horizontal and diagonal directions of the square lattice, and by time reversal. Hence, if there is a monopole at $\mathbf{\Phi}_i = (\Phi_{1,i}, \Phi_{2,i}, \Phi_{3,i})$, it must be accompanied by monopoles at $(\pm \Phi_{1,i}, \pm \Phi_{2,i}, \pm \Phi_{3,i})$. A monopole at the origin, $\mathbf{\Phi}_i  = 0$, does not have symmetry relatives. Note that the components of $\mathcal{B}$ transform in the same way as those of $\mathbf{\Phi}$ under the point group operations and time reversal (see Section \ref{sec:symmetry} of this Supplement).

Imagine deforming our theory, e.g., by changing the microscopic Hamiltonian, while maintaining all the symmetries of the problem. An isolated monopole of $\mathcal{B}$ at $\mathbf{\Phi}_0 = 0$ with strength $q_0$ can change into a monopole of strength $q_0 - 2$, by splitting into three monopoles along any of the three $\Phi$ axes, of strengths $1$, $q_0-2$, and $1$, respectively. Importantly, the \emph{parity} of $q_0$ is invariant under such deformations. We can thus classify the allowed Berry phase terms according to the parity of $q_0$ of the monopole at the origin.   

To understand the two classes of Berry phase terms, we examine their microscopic origin. Consider the low-lying states in the Hilbert space of a single unit cell. 
These states transform under the point group symmetry transformations as one of the irreducible representations of D$_{4h}$. If the low-energy states form a two-dimensional representation (E$_g$ or E$_u$), then these states are doubly degenerate. In this situation, all the low-energy states of an $L\times L$ unit cell system (with odd $L$) form doubly degenerate multiplets. This can be seen by the fact that, acting on a single site, the mirror transformations along the diagonal and the horizontal directions anti-commute with each other: $M_d M_h = - M_h M_d$. This property is preserved for any finite-size lattice with an odd number of sites that is symmetric under the point group. 

If a symmetry breaking field is applied (either of the $\Phi_{1,2,3}$ character), this degeneracy is split. In the path integral formulation, a monopole at $\mathbf{\Phi}=0$ of strength $q_0=1$ encodes the double degeneracy at the origin of $\mathbf{\Phi}$ space. This is similar to the Berry phase term in a coherent-state path integral of a single spin-$\frac{1}{2}$ particle. For example, in the model consisting of a pseudospin arising from an $E_g$ doublet, $M_h M_d =  - M_d M_h$ when acting on a single unit cell (or any finite lattice of size $L\times L$ unit cells with odd $L$), and thus $q_0=1$. 

Note that this argument relies on having a well-defined transformation law of the low-lying states of each unit cell. It does not apply if the local Hilbert space contains multiplets transforming as different irreducible representations of the point group. In particular, this complicates the analysis if the system is metallic, since adding a single electron to a unit cell can change the transformation properties of the low-lying states.

To understand the effect of the Berry-phase term, let us consider the case in which there is a field conjugate to one component of the three, so we can count on this component always having a substantial magnitude.  For convenience, let us consider the problem in the presence of a transverse field $h_1>0$, 
so that in doing the path integral we can always assume that $\Phi_1$ is substantial. For an appropriate gauge choice and rescaling of the components of $\mathbf{\Phi}$ such that 
\begin{eqnarray}
{\bm A}({\bm \Phi}) = q_0 \left[\frac {\Phi-\Phi_1}{4\pi \Phi}\right] \left[\frac {\Phi_2{\bm e}_3 - \Phi_3 {\bm e}_2}{\Phi_2^2+\Phi_3^2}\right],
\end{eqnarray}
where $\Phi=\sqrt{\Phi_1^2+\Phi_2^2+\Phi_3^2}$.
 This gauge corresponds to having a quantized flux coming in through an infinitely narrow solenoid 
 along the $-{\bm e}_1$ axis and then having the magnetic flux spread out from the origin where it terminates. This corresponds to the $\mathcal{B}$ field of a nematic monopole, as described above. Notice that 
 for $\Phi_{2,3} \ll \Phi$
\begin{equation}
{\bm A}({\bm \Phi}) = q_0 \left[\frac {1}{8\pi {\Phi_1^2}}\right] \left[ {\Phi_2{\bm e}_3 - \Phi_3 {\bm e}_2}\right] \left[1 + {\cal O} \left(\frac {\Phi_2^2+\Phi_3^2}{\Phi^2}\right)\right].
\end{equation}

We are interested in the behavior of $\Phi_3$ in the disordered state or in the ordered state not too far from the QCP, so that the condition $|\Phi_3| \ll |\Phi_1|$ remains valid. We can safely approximate $\Phi_1$ by a non-zero constant value, $\bar \Phi_1$, which is an increasing function of the transverse field, $h_1$. Assuming that we are not near a time reversal broken phase, we can integrate out $\Phi_2$, which, since its fluctuations are small, can be treated in a Gaussian approximation. The resulting effective action for $\Phi_3$ is
\begin{equation}
S^{\mathrm{eff}}[\Phi_3] = S_0[\Phi_3] +\frac {q_0^2\chi_2} {128{|\bar \Phi_1|^4}} \sum_{\vec R} \int_0^\beta \ d\tau\ {|\dot\Phi_3|^2}
\end{equation}
where $\chi_2$ is the susceptibility with respect to the other transverse field, $h_2$, and $S_0$ has all the $\Phi_3$ dependent terms from $S_{LGW}$.

In short, there is an additive contribution to the effective mass for $\Phi_3$ that is proportional to $q_0^2$, i.e. it comes from the quantum character of the fluctuations.  This term is, in turn, a strongly decreasing function of $|\bar\Phi_1|$;  an increasing transverse field thus causes a decrease in the effective mass, thereby increasing the quantum fluctuations of $\Phi_3$.

Before concluding this section, let us comment on the case of a metallic system undergoing a ferroquadrupolar transition. As mentioned above, in this case, the effective action for the order parameter field $\mathbf{\Phi}$ is complicated by the presence of non-local terms. Therefore, strictly speaking, the analysis presented in this section does not apply. However, if our system is composed of localized quadrupoles coupled to itinerant electrons, as in the $4f$ systems TmAg$_2$ and TmAu$_2$ discussed in the main text, we may imagine artificially setting the coupling between the two sets of degrees of freedom to zero, while keeping the exchange coupling between the local quadrupoles finite. Then, applying a diagonal strain tunes through a ferroquadrupolar QCP. Turning the coupling between the itinerant electrons and the local quadrupoles back on, we expect the QCP to survive over a finite range of coupling. Therefore, one may expect that a diagonal strain can tune through a QCP in the metallic case as well, although the properties of the QCP are very different from the case of an insulator.


\section{Transformation of $\mathbf{\Phi}$ under symmetries}

\label{sec:symmetry}

The symmetries of the components of $\mathbf{\Phi}$ under the horizontal
and diagonal mirror symmetries, $M_{h}$ and $M_{d}$, and under time
reversal, $\mathcal{T}$ are summarized in Table \ref{tab:Sym}. We
can construct the combination of the symmetry operations, $\tilde{M}_{h}=M_{h}\mathcal{T}$,
$\tilde{M}_{d}=M_{d}\mathcal{T}$, and $\mathcal{T}$, such that under
each of these symmetries, one component of $\mathbf{\Phi}$ is odd,
while the other two are even. Thus, these symmetries act as reflections
in $\mathbf{\Phi}$ space about the $\Phi_{2}-\Phi_{3}$, $\Phi_{3}-\Phi_{1}$,
and $\Phi_{1}-\Phi_{2}$ planes, respectively.

The components of $\mathbf{A}(\mathbf{\Phi})$ transform under $M_{d,h}$
in the same way as those of $\mathbf{\Phi}$. Under $\mathcal{T}$,
they transform oppositely to the components of $\mathbf{\Phi}$. From
this, we can derive the transformation law of the Berry curvature.
$\mathcal{B}_{a}=\varepsilon_{abc}\partial A_{b}/\partial\Phi_{c}$
transforms as $A_{a}$ (and $\Phi_{a}$) under the point group operations,
and in the same way as $\Phi_{a}$ (i.e., oppositely to $A_{a}$)
under time reversal. For example, consider $\mathcal{B}_{3}=\partial A_{1}/\partial\Phi_{2}-\partial A_{2}/\partial\Phi_{1}$;$A_{1}$
and $\Phi_{2}$ are both $\mathcal{T}$ odd, while $A_{2}$ and $\Phi_{1}$
are both even, so $\mathcal{B}_{3}$ is $\mathcal{T}$ even, as is
$\Phi_{3}$.

\begin{table}[h]
\caption{Transformation laws of $(\Phi_{1},\Phi_{2},\Phi_{3})$ under point
group transformations and time reversal.\label{tab:Sym}}

\begin{tabular}{|c|c|c|c|}
\hline 
Symmetry  & $\Phi_{1}$  & $\Phi_{2}$  & $\Phi_{3}$\tabularnewline
\hline 
\hline 
$M_{h}$  & $-$  & $-$  & $+$\tabularnewline
\hline 
$M_{d}$  & $+$  & $-$  & $-$\tabularnewline
\hline 
$R_{\pi/2}=M_{h}M_{d}$  & $-$  & $+$  & $-$\tabularnewline
\hline 
$\mathcal{T}$  & $+$  & $-$  & $+$\tabularnewline
\hline 
\end{tabular}
\end{table}

\section{Mean field theory in a metallic system - dichotomy between 1 and
2 band models}

\subsection{The models}

Let us try to examine the dichotomy between single band and multi-band
versions of the transverse field Ising model, at the level of a mean-field
field theory. We consider starting from the 1- and 2- band Hamiltonians:
\begin{align}
H_{1} & =\sum_{\k}\epsilon_{1\k}c_{\k}^{\dag}c_{\k}+H_{1\text{int}}\\
H_{2} & =\sum_{\k}\left(c_{x\k}^{\dag}\,\,c_{y\k}^{\dag}\right)\,\begin{pmatrix}\epsilon_{2x\k} & 0\\
0 & \epsilon_{2y\k}
\end{pmatrix}\begin{pmatrix}c_{x\k}\\
c_{y\k}
\end{pmatrix}+H_{2\text{int}}
\end{align}
In the second Hamiltonian, the orbital degrees of freedom have $x$
and $y$ symmetries (e.g. $d_{xz}$ and $d_{yz}$ orbitals). We imagine
that the interaction terms $H_{1\text{int}}$ and $H_{2\text{int}}$
serve to drive both systems into a nematic phase, with $x^{2}-y^{2}$
symmetry.

For the single band model, we choose the dispersion to be that of
a square lattice, nearest neighbor tight binding model, which in the
presence of the $xy$ strain has the form : 
\begin{align}
\epsilon_{1\k} & =-2t(\cos{k_{x}}+\cos{k_{y}})-4t^{\prime}\cos{k_{x}}\cos{k_{y}} +\varepsilon_{xy}\sin{k_{x}}\sin{k_{y}}
\end{align}
(we will set $t=1$), while for the two band model, symmetry under
$x\rightarrow y$ means we can choose: 
\begin{align}
\epsilon_{2x\k} & =-2t_{a}\cos{k_{x}}-2t_{b}\cos{k_{y}}=-2t_{1}s_{\k}-2t_{2}d_{\k}\\
\epsilon_{2y\k} & =-2t_{b}\cos{k_{x}}-2t_{a}\cos{k_{y}}=-2t_{1}s_{\k}+2t_{2}d_{\k}
\end{align}
where I have defined the $s$-wave and $d$-wave form factors 
\begin{align}
s_{\k} & =\cos{k_{x}}+\cos{k_{y}}\\
d_{\k} & =\cos{k_{x}}-\cos{k_{y}}.
\end{align}
Here, the $xy$ strain term has the form $\varepsilon_{xy}\Psi_{\k}^{\dag}\sigma_{x}\Psi_{\k}$
where $\Psi_{\k}^{\dag}=(c_{x\k}^{\dag}\,\,c_{y\k}^{\dag})$ is a
two component spinor as usual.

\subsection{Self consistency equations}

To derive the appropriate mean-field self consistency equations, we
assume that the dominant interaction in the case of the 1 band model
is a forward scattering type interaction: 
\begin{align}
H_{1\text{int}}=\frac{-V}{N}\sum_{\k,\k^{\prime}}d_{\k}d_{\k^{\prime}}\,\,c_{\k}^{\dag}c_{\k}c_{\kp}^{\dag}c_{\kp}
\end{align}
where $N$ is the system size. Meanwhile in the two band model, the
dominant interaction is 
\begin{align}
H_{2\text{int}}=-\frac{V}{N}\sum_{\k,\k^{\prime}}\left(\Psi_{\k}^{\dag}\sigma_{z}\Psi_{\k}\right)\left(\Psi_{\kp}^{\dag}\sigma_{z}\Psi_{\kp}\right)
\end{align}
We will introduce mean field order parameters in both models, of the
form 
\begin{align}
\phi_{1} & =\frac{V}{N}\sum_{\k}d_{\k}\langle c_{\k}^{\dag}c_{\k}\rangle_{\text{MF}}\\
\phi_{2} & =\frac{V}{N}\sum_{\k}\langle\Psi_{\k}^{\dag}\sigma_{z}\Psi_{\k}\rangle_{\text{MF}}
\end{align}
where expectation values are taken in the quadratic Hamiltonians:
\begin{align}
H_{1,\text{MF}} & =\sum_{\k}\varepsilon_{1\k,\text{MF}}c_{\k}^{\dag}c_{\k} =\sum_{\k}\left(-2s_{\k}-2\phi_{1}\,d_{\k}+4t^{\prime}\cos{k_{x}}\cos{k_{y}}+\varepsilon_{xy}\sin{k_{x}}\sin{k_{y}}\right)c_{\k}^{\dag}c_{\k}
\end{align}
and 
\begin{align}
H_{2,\text{MF}} & =\sum_{\k}\Psi_{\k}^{\dag}\left[-2t_{1}s_{\k}I-2(t_{2}d_{\k}+\phi_{2})\sigma_{z}+\varepsilon_{xy}\sigma_{x}\right]\Psi_{\k}
\end{align}
The self consistency equation for the 1 band model takes the form:
\begin{equation}
\phi_{1}=\frac{V}{N}\sum_{\k}\left(\cos{k_{x}}-\cos{k_{y}}\right)f\left(\varepsilon_{1\k,\text{MF}}\right)
\end{equation}
where $f$ is the Fermi distribution function, while for the two band model it is: 
\begin{equation}
\phi_{2}=\frac{V}{N}\sum_{\k}\frac{2\left(\phi_{2}+t_{2}d_{\k}\right)}{\sqrt{4(t_{2}d_{\k}+\phi_{2})^{2}+\varepsilon_{xy}^{2}}}\left[f(\varepsilon_{-\k})-f(\varepsilon_{+\k})\right],
\end{equation}
where 
\begin{align}
\varepsilon_{\pm\k}=-2t_{1}s_{\k}\pm\sqrt{4(t_{2}d_{\k}+\phi_{2})^{2}+\varepsilon_{xy}^{2}}
\end{align}

\subsection{Results}

\begin{figure}[t]
\includegraphics[width=0.8\textwidth]{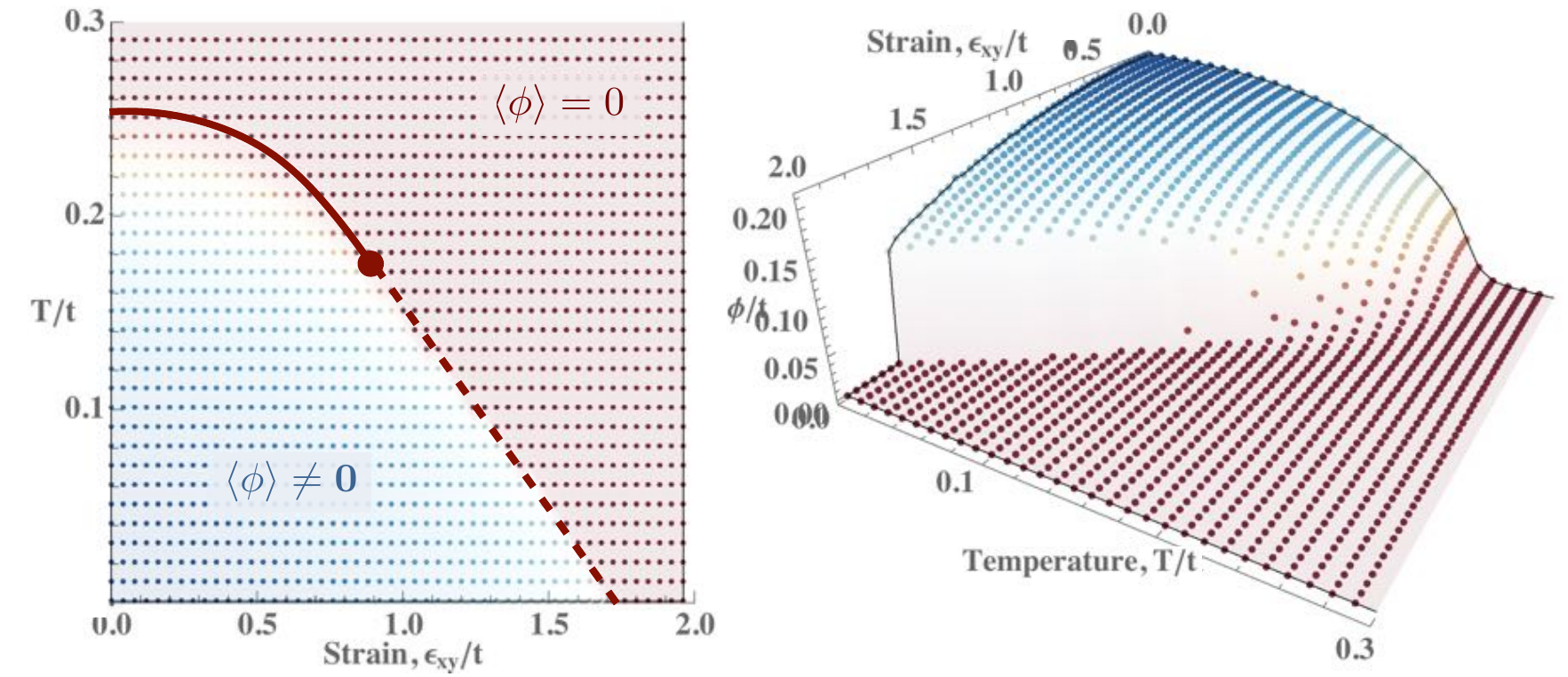} \caption{A summary of the phase diagram for a one band model, with bandstructure
parameters $t^{\prime}=0.05t$, $\mu=-0.2t$ and an interaction strength
$V=1.3t$. We find that the quantum phase transition is generically
first order (dashed lines) for the single band model - a continuous
transition (solid line) seems to require excessive fine tuning. (The
colors used indicate the strength of the nematic order parameter blue
corresponds to non-zero $\phi$, dark red is $\phi=0$.) }
\label{fig:mft1band} 
\end{figure}

\begin{figure}
\includegraphics[width=0.8\textwidth]{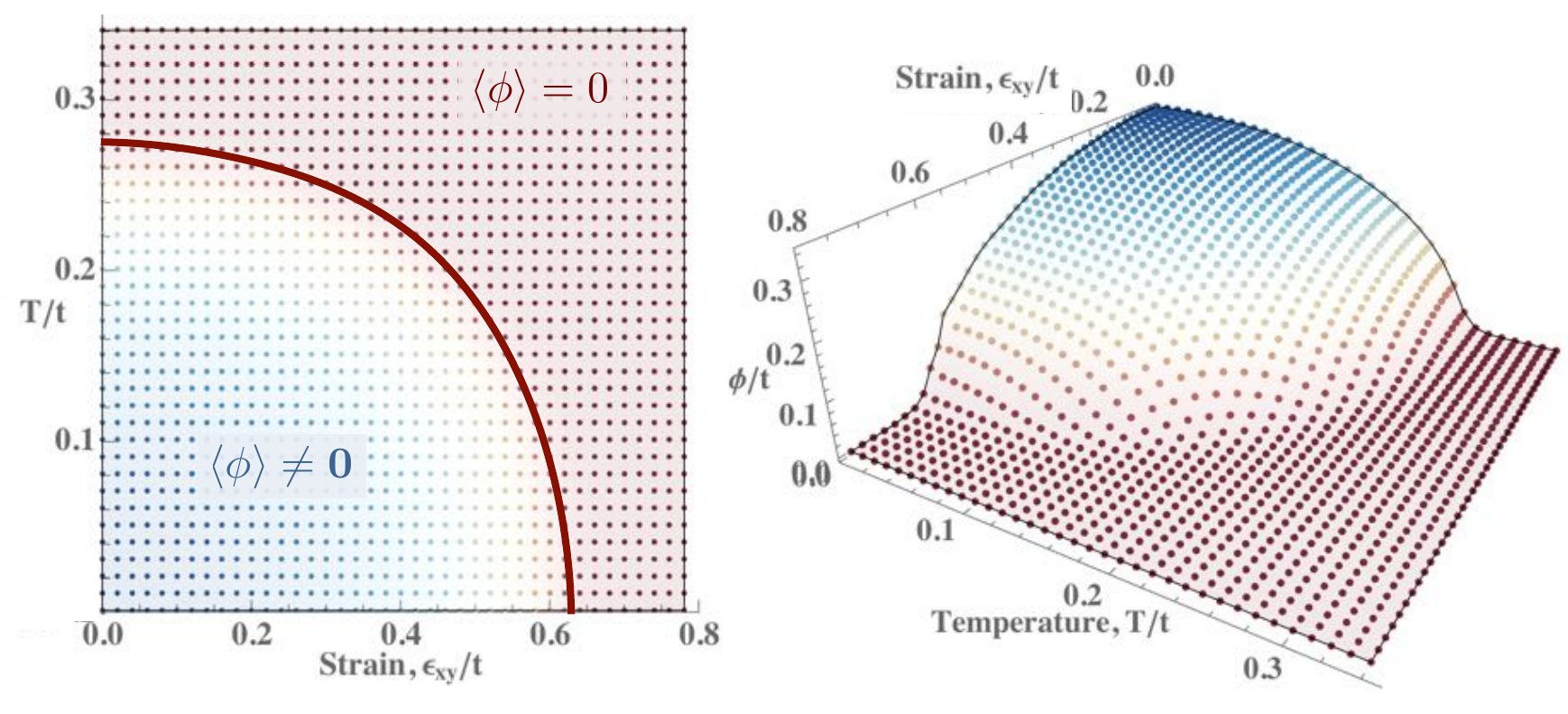} \caption{A summary of the phase diagram for a two band model, with bandstructure
parameters $t_{2}=0.05t$, $\mu=0$ and an interaction strength $V=1.2t$.
So long as the hopping anisotropy $t_{2}$ is small, it is relatively
`easy' to find a continuous quantum phase transtion. (The colors used
indicate the strength of the nematic order parameter : blue is a well
developed value, dark red is $\phi=0$.) }
\label{fig:mft2band} 
\end{figure}

We now numerically solve these self consistency equations for a model
with a single orbital per site (Fig ~\ref{fig:mft1band}), and then
for a 2 orbital model (Fig. ~\ref{fig:mft2band}). We have solved
the mean field equations on a 2 dimensional lattice with $256\times256$
points by iterating the self consistency equations until convergence
is achieved.

Generically we find that for the single band metal, the nematic phase
terminates at a first order quantum phase transition. This is in fact
consistent with previous studies \cite{hae-youngPRB2004,vadimPRB2005}

On the other hand, continuous zero temperature transitions (i.e. quantum
critical points) are possible for the two band model (Fig. ~\ref{fig:mft2band}),
provided the hopping anisotropy parameter $t_{2}$ is not too large.

\FloatBarrier
\section{Quadrupolar order in $4f$ intermetallics}
As we discussed in the main text, the $4f$ intermetallic compounds TmAg$_2$ and TmAu$_2$ provide a specific realization of Ising nematic order arising from the splitting of a non-Kramers doublet.  Here, we show how crystal field effects give rise to an $E_g$ doublet as the lowest energy eigenstates, and then discuss how Ising nematic (ferroquadrupolar) order arises using a mean field treatment of the quadrupole-quadrupole interactions.

\subsection{Crystal field effects in a $4f$ system: how a doublet arises from a $J=6$ state.}

In the rare earth intermetallic compounds TmAg$_2$ and TmAu$_2$, the Tm ion takes the Tm$^{3+}$ ($4f^{12}$) state. From Hund's rules the $4f$ electronic orbitals are filled such that $L=5$ and $S=1$, giving the electronic multiplet a total angular momentum state of $J=6$. Thus for a local Tm site the spherical harmonics $Y^{m}_6$ form the natural basis to construct wavefunctions. The surrounding Au ions create a crystalline electric potential which obeys the tetragonal point group symmetry $ D_{4h}$, and so the degeneracy of the 13 $Y^{m}_6$ states is removed . The nature of the resulting eigenstates (but not their relative energies) can be inferred purely from symmetry arguments. Here we demonstrate this analysis explicitly, following closely the manipulations described in \cite{dresselhaus2007group}.
\begin{table*}
\begin{center}
\begin{tabular*}{\textwidth}{c c | @{\extracolsep{\fill}} c  c  c  c  c  c  c  c  c  c  c }
\hline 
Irrep & BSW notation\footnote{due to Bouckaert, Smoluchowski \& Wigner (1936)} & $E$ & $2C_4$ & $C_2$ & $2C'_2$ & $2C''_2$ & $i$ & $2S_4$ & $\sigma_h$ & $2\sigma_v$ & $2\sigma_d$\\
\hline 
$A_{1g}$ & $\Gamma^{+}_1$ & 1 & 1 & 1 & 1 & 1 & 1 & 1 & 1 & 1 & 1 &   \\
$A_{2g}$ & $\Gamma^{+}_2$ &  1 & 1 & 1 & -1 & -1 & 1 & 1 & 1 & -1 & -1 &  \\
$B_{1g}$ &  $\Gamma^{+}_3$ & 1 & -1 & 1 & 1 & -1 & 1 & -1 & 1 & 1 & -1 &   \\
$B_{2g}$ & $\Gamma^{+}_4$ &  1 & -1 & 1 & -1 & 1 & 1 & -1 & 1 & -1 & 1 &   \\
$E_{g}$ &  $\Gamma^{+}_5$ & 2 & 0 & -2 & 0 & 0 & 2 & 0 & -2 & 0 & 0 & \\
$A_{1u}$ &  $\Gamma^{-}_1$ & 1 & 1 & 1 & 1 & 1 & -1 & -1 & -1 & -1 & -1 &  \\
$A_{2u}$ &  $\Gamma^{-}_2$ & 1 & 1 & 1 & -1 & -1 & -1 & -1 & -1 & 1 & 1 & \\
$B_{1u}$ & $\Gamma^{-}_2$ &  1 & -1 & 1 & 1 & -1 & -1 & 1 & -1 & -1 & 1 &  \\
$B_{2u}$ & $ \Gamma^{-}_4$ & 1 & -1 & 1 & -1 & 1 & -1 & 1 & -1 & 1 & -1 & \\
$E_{u}$ & $\Gamma^{-}_5$ &  2 & 0 & -2 & 0 & 0 & -2 & 0 & 2 & 0 & 0 &  \\
\hline
\rule{0pt}{3ex}
$\Gamma^{J=6}$ && 13 & -1 & 1 & 1 & 1 & 13 & -1 & 1 & 1 & 1 &  \\
\rule{0pt}{3ex}
$\Gamma^{J}$ & &$(2J+1) $&$( -1)^{\text{floor}(J/2)}$ & $(-1)^{J}$ &  $(-1)^{J}$  &  $(-1)^{J}$  &$ (-1)^{J}(2J+1)$&$ ( -1)^{J+\text{floor}(J/2)}$ & 1 & 1 & 1 &  \\
\hline
\end{tabular*}
\caption{Character Table for $D_{4h}$, with the reducible representations $\Gamma^{6}$ and for general $J$, $\Gamma^{J}$ shown at the bottom. Determining the crystal field splitting of the $J=6$ Hund's rule ground state amounts to determining how the $\Gamma^{6}$ representation (which is \textit{reducible} in $D_{4h}$) is decomposed into irreducible representations of $D_{4h}$. }
\end{center}
\label{table:mulliken}
\end{table*}

The general goal is to examine how the $\Gamma^{J=6}$ representation of $SO(3)$ is reduced in the tetragonal environment (with point group $D_{4h}$), i.e. we would like to find the coefficients $a_i$ in
\begin{align}
\Gamma^{J=6} = \sum_i a_i \Gamma^{}_{i}
\end{align}
where $\Gamma^{}_{i}$ are the irreducible representations of $D_{4h}$. To do this, we must follow these steps:
\begin{itemize}
\item First, calculate the character $\chi^{(\text{reducible})}(\mathcal{C}_k)$ of the (reducible) $\Gamma^{J=6}$ representation under the group operations $\mathcal{C}_k$ of $D_{4h}$. Using the 13 $Y^m_6$ spherical harmonics as a basis, the $\chi^{(\text{reducible})}(\mathcal{C}_k)$ is the trace of the matrix that corresponds to the group element $\mathcal{C}_k$. 
\item We then use the orthogonality of characters to obtain the coefficients $a_j$:
\begin{align*} 
a_j = \frac{1}{h}\sum_k N_k \chi^{(\text{reducible})}(\mathcal{C}_k) \chi^{(\Gamma_n)}(\mathcal{C}_k)
\end{align*}
where $h = 16$ is the order of the group $D_{4h}$, $N_k$ is the number of elements in class $k$, and $\chi^{(\Gamma_n)}(\mathcal{C}_k)$ is the character of the \textit{irreducible} representation $k$ in $D_{4h}$.
\end{itemize}
For example, it can be shown that under a rotation about the $z$ axis by $\alpha$, the spherical harmonic $Y^{m}_{J}$ is transformed as
\begin{align}
Y^{m}_{J}(\theta,\phi+\alpha) = e^{im\alpha}Y^{m}_{J}(\theta,\phi),
\end{align}
so that, under the group element corresponding to four fold rotations ($C_4$),  the character is 
\begin{align*}
\chi^{(\text{reducible})}(C_4)  = \sum^{6}_{m=-6} e^{i\frac{\pi m}{2}} = -1
\end{align*}
The characters of the remaining symmetry operations in $D_{4h}$ have been appended to the character table (Table S2), both for arbitrary $J$, and then specifically for $J=6$. Having written the reducible representation $\Gamma^{J=6}$, we can now reduce it on the irreducible representations of $D_{4h}$, using the orthogonality of characters. This gives
\begin{align}
\Gamma^{J=6} = 2A_{1g} \oplus A_{2g} \oplus 2B_{1g} \oplus 2B_{2g} \oplus 3E_g
\end{align}
This implies that the CEF Hamiltonian breaks the degeneracy of the 13 $J=6$ states in such a way that there are 3 $E_{g}$ doublets, and 7 singlets. Two of the singlets have wavefunctions which have  $ A_{1g}$  symmetry, one that has $A_{2g}$ symmetry, two have $ B_{1g}$ symmetry, and two have $B_{2g}$ symmetry. We stress that group theory does not determine the relative energies of these representations. For TmAg$_2$, the ground state is determined by the eigenvalues of the CEF Hamiltonian exhibiting the $D_{4h}$ point group symmetry surrounding the Tm sites. This Hamiltonian can be written as a sum of the Stevens operators $\hat{O}^{m}_l$ :
\begin{align}
\hat{H}_{CEF}=B^0_2\hat{O}^0_2+B^0_4\hat{O}^0_4+B^4_4\hat{O}^4_4+B^0_6\hat{O}^0_6+B^4_6\hat{O}^4_6
\end{align}
The coefficients $B^m_l$ have been experimentally inferred through inelastic neutron scattering and magnetic susceptibility to be:
 $B^0_2=0.0909$K, $ B^0_4=0.3265$mK, $B^4_4=-64.5$mK, $B^0_6=0.0897$mK, $B^4_6=-1.30$mK \cite{MorinTMAg2PRB93}. Diagonalizing this CEF Hamiltonian gives the energy spectrum displayed below.
\begin{figure}[h]
\centering
\includegraphics[width=0.9\textwidth]{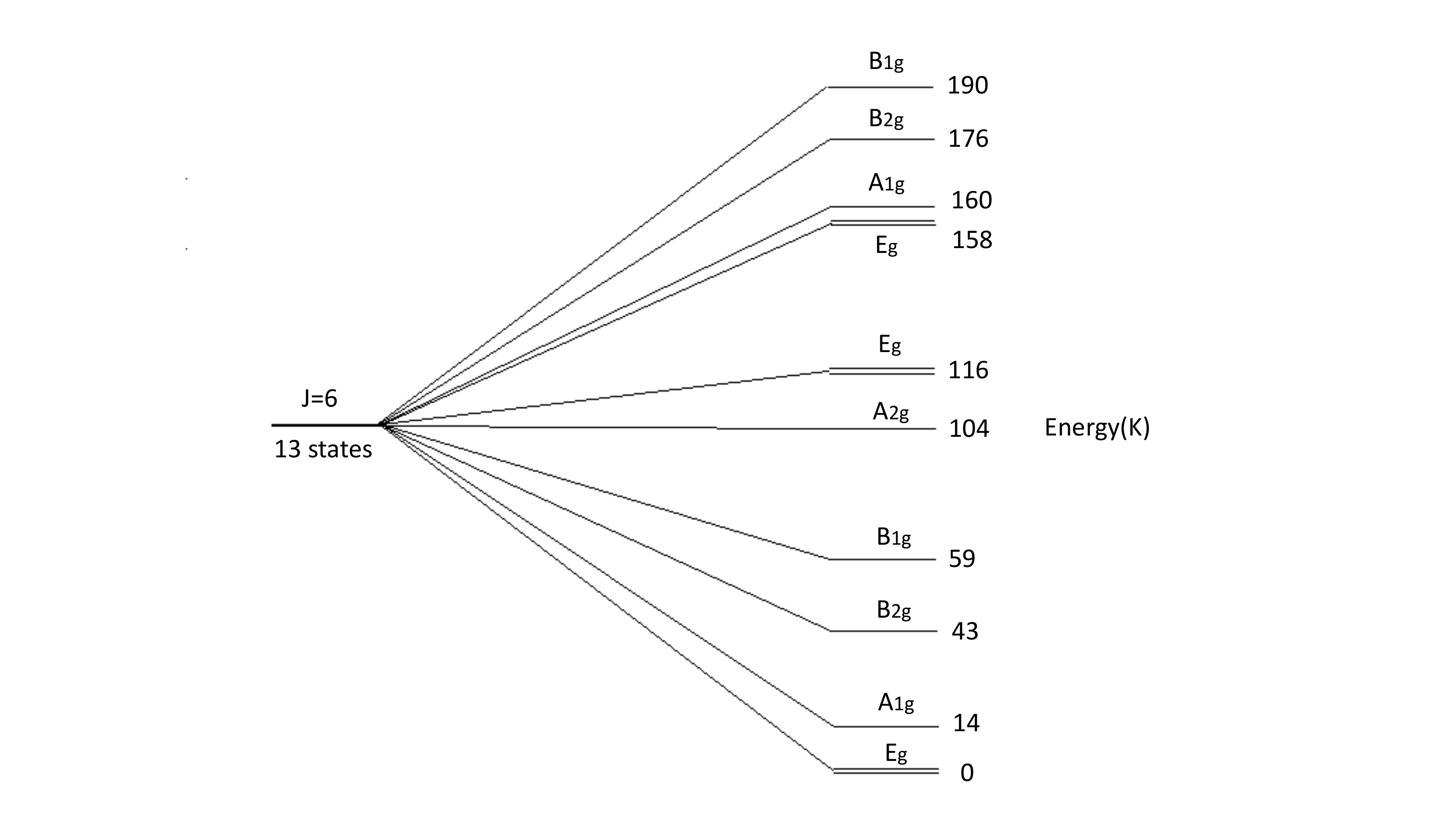}
\caption{The Energy levels of TmAg$_2$ when the local $J=6$ state is reduced in the $D_{4h}$ Crystalline Electric Field}
\end{figure}
\FloatBarrier
The ground state of TmAg$_2$ is an $E_g$ doublet with the closest excited state, an $A_{1g}$ singlet, at 14K. These states determine much of its low temperature behavior.

\subsection{Projecting to the non Kramer's $E_g$ doublet}
The basis vectors corresponding to different irreducible representations can be found by use of the projection operator:
\begin{align}
\hat{P}^{(\Gamma_{n})}=\frac{l_n}{h}\sum_{\mathcal{C}_{k}}\chi^{(\Gamma_n)}\left(\mathcal{C}_k\right)\hat{P}_{\mathcal{C}_{k}},
\label{eq:projectionOp}
\end{align}
where $\Gamma_{n}$ is the representation of interest, $ l_n$ is the dimension of the representation, $h$ is the order of the group, $\chi^{(\Gamma_n)}\left(\mathcal{C}_k\right)$ is the character of $\Gamma_n$ for group element $\mathcal{C}_k$, and $\hat{P}_{\mathcal{C}_k}$ is corresponding operator. When the projection operator is applied to an arbitrary function, it projects a linear combination of the basis vectors $\ket{\Gamma_{n}k}$ for that particular irreducible representation.
\begin{align}
\hat{P}^{(\Gamma_{n})}F=\sum_{k}{A_{k}^{(\Gamma_{n})}\ket{\Gamma_{n}k}}
\end{align}
For the case of the $E_{g}$ doublet, $l_{n} =2$, $h=16$ for $D_{4h}$, and we can look at the character table above to find the traces of the matrices. Assuming an arbitrary function consisting of all 13 degenerate states:
\begin{align}
\hat{P}^{(\Gamma_{5})}F&=\frac{1}{8}(2E-2C_{2}+2i-2{\sigma_h})\begin{bmatrix} f_{1}\ket{6}\\f_{2}\ket{5} \\.\\.\\f_{13}\ket{-6} \end{bmatrix}
=A_{k}^{(\Gamma_{5})}\ket{\Gamma_{5}k}\\
\hat{P}^{(\Gamma_{5})}F &=\begin{bmatrix} 0&.&.&.&&\hdots \\  .&1& & & & \\ .&&0&& &\\.& &&1&  &\\ & && &0 & \\\vdots& && &&\ddots\end{bmatrix}\begin{bmatrix} f_{1}\ket{6}\\f_{2}\ket{5} \\.\\.\\f_{13}\ket{-6} \end{bmatrix}
=A_{k}^{(\Gamma_{5})}\ket{\Gamma_{5}k}
\end{align}
Thus the only possible states that can compose the $ E_{g} $ doublet states are the odd states. Furthermore, given the behavior of spherical harmonics under the rotation $C_4$ one can show that only states that combine angular momenta that differ by $\Delta m = 4$ are allowed. To see this, consider the action of $\hat{C}_4$ on a state $\ket{\psi} = A \ket{m_1} + B \ket{m_2}$. We have $\hat{C}_4 \psi = e^{im_1\pi/2}\left( A\ket{m_1} + e^{i(m_2-m_1)\pi/2}\ket{m_2}\right)$. For $\psi$ to be an eigenstate, we must have $(m_2 - m_1)\pi/2 = 2n\pi$, i.e. $m_2 - m_1 = 4n$. This, in addition to the fact that the doublet must preserve time reversal symmetry forces us to conclude that the $E_{g}$ representation can be expressed in the spherical harmonic basis as
\begin{align} 
\ket{\psi^{\Gamma_{5}}_1}&=e\ket{5}+f\ket{1}+g\ket{-3}\\
\ket{\psi^{\Gamma_{5}}_2}&=e\ket{-5}+f\ket{-1}+g\ket{3}
\end{align}
Now that we know the form of the $E_g$ doublet, let us examine the effects of applying a magnetic field as well as applying strains. In a vector space consisting of basis vectors $\ket{\psi^{\Gamma_{5}}_1}$ and $ \ket{\psi^{\Gamma_{5}}_2}$, a magnetic field $H_z$ applied in the $z$-direction measures the $z$ component of angular momentum for each state, and thus is proportional to the operator $J_z$:
\begin{align}
\bra{\psi^{\Gamma_{5}}_i}\hat{H}_{z}\ket{\psi^{\Gamma_{5}}_j}=\begin{bmatrix}C &0\\0&-C \end{bmatrix} = C \,\hat{\sigma}_z,
\end{align}
where $C=5|e|^2+|f|^2-3|g|^2$. If we were to apply the $B_{1g}$ strain $\varepsilon_{xx}-\varepsilon_{yy}$ the corresponding quadrupole operator $\hat{O}^2_2=J_x^2-J_y^2$ takes the form:
\begin{align}
\bra{\psi^{\Gamma_{5}}_i}\hat{O}^2_2\ket{\psi^{\Gamma_{5}}_j}=\begin{bmatrix}0 &D\\D&0 \end{bmatrix}= D \,\hat{\sigma}_x,
\end{align}
where $D=21f^2+12\sqrt{10}fg+2\sqrt{165}eg$ assuming real coefficients. Finally, the operator for $B_{2g}$ nematic strain $\varepsilon_{xy}$ is the quadrupole operator $\hat{P}_{xy}=\frac{1}{2}(J_{x}J_{y}+J_{y}J_{x})$ which takes the form: 
\begin{align}
\bra{\psi^{\Gamma_{5}}_i}\hat{P}_{xy}\ket{\psi^{\Gamma_{5}}_j}=\begin{bmatrix}0 &-iE\\iE&0 \end{bmatrix} = E\, \hat{\sigma}_y,
\end{align}
where $E=\frac{21}{2}f^2-6\sqrt{10}fg+\sqrt{165}eg$.

These operators can clearly be identified with the Pauli matrices (Pseudospin operators). Note that a change of basis will rotate these operators into the canonical orbital basis that was used in the main text. We have therefore shown how projecting onto the ground state $E_g$ doublet of this system gives rise to the pseudo-spin 1/2 operators.

\subsection{Quadrupole-Quadrupole interactions and the spin 1/2 transverse field Ising model}
\label{sec:effectiveH}
As discussed in the main text, a generalized version of RKKY interactions \cite{Schmitt1985,MorinSchmitt88,SchmittQuadrupolar90, Morin1990Quadrupolar} gives rise to Quadrupole-Quadrupole (QQ) interactions. In the presence of a $B_{2g}$ strain, the full Hamiltonian of the $J=6$ system is then:
\begin{align}
H_{J=6} &= \sum_{i} \hat{H}_{CEF} - \sum_{i,j} J_{ij} \hat{O}^{2}_{2,i} \hat{O}^{2}_{2,j} + B_{B{2g}}\sum_{i} \varepsilon_{xy}\hat{P}_{xy,i}
\label{eq:fullh}
\end{align}
Here, $J_{ij}$ is the strength of (attractive) QQ interactions (including strain renormalizations) between sites $i$ and $j$, $\hat{H}_{CEF}$ is the crystal field Hamiltonian outlined in the previous sections, and $B_{B2g}$ is the strength of the coupling to external strains $\varepsilon_{xy}$. Here, $\hat{O}^2_2$ and $P_{xy}$ are the full $13\times 13$ matrices representing Stevens operators. In Section \ref{sec:mfteq} we describe our mean field treatment of this full 
Hamiltonian. 

Before doing so, let us note that the spin 1/2 version of this Hamiltonian may be obtained by projecting these operators onto the ground state $E_g$ doublet of $\hat{H}_{CEF}$. Applying this projecting operator, we find the projected version of this Hamiltonian to be 
\begin{align}
H_{\text{proj}} = -\sum_{i,j}\tilde{J}_{ij} \hat{\sigma}_{x,i}\hat{\sigma}_{x,j} + \tilde{B}\sum_{i} \varepsilon_{xy}\sigma_{y,i}
\end{align}
where $\tilde{J}$ and $\tilde{B}$ are proportional to the original constants $J_{ij}$ and $B_{B2g}$ of Eq.~\ref{eq:fullh}. Thus, our mapping to the spin $1/2$ transverse field Ising model is complete. 

\subsection{Mean field equations}
\label{sec:mfteq}
\begin{figure}
\includegraphics[width=0.48\textwidth]{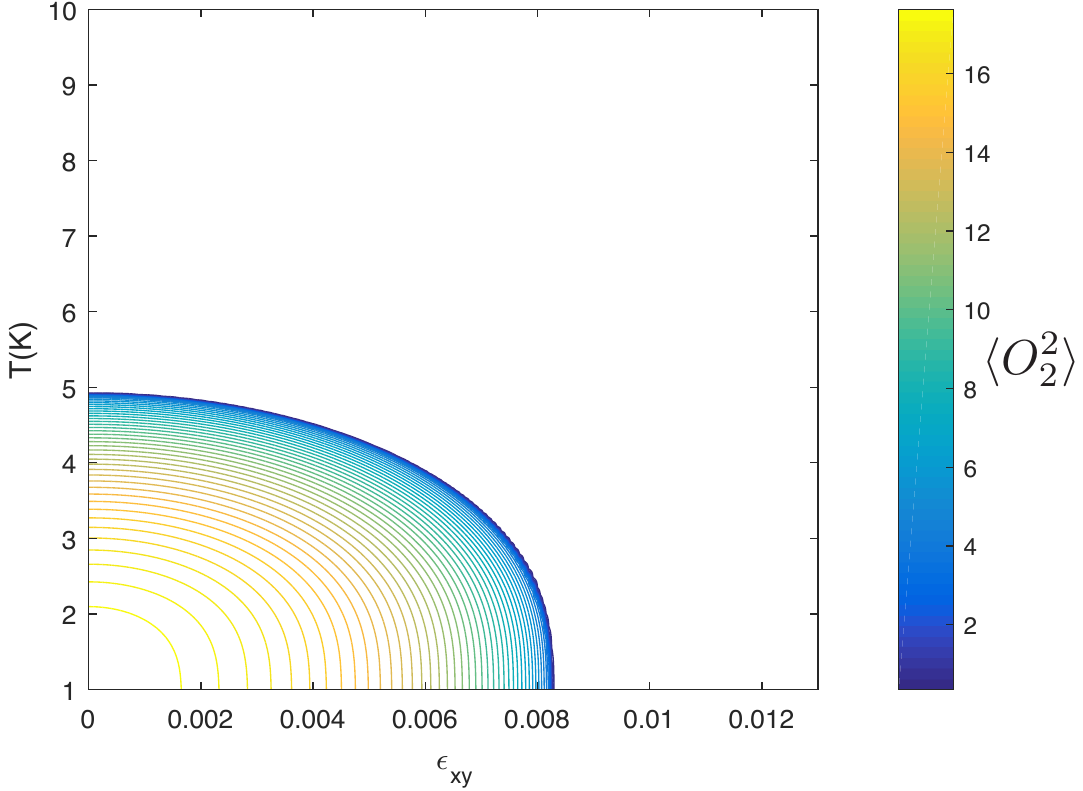}\quad
 \caption{The phase diagram obtained for the mean field calculation of the magnitude of ferroquadrupolar order as a function of temperature and  transverse strain $\varepsilon_{xy}$ in TmAg$_2$. The parameters in the mean field Hamiltonian are entirely determined by experiment\cite{MorinTMAg2PRB93}, except for $B_{xy}$ which was assumed to be $100$K (a similar value to that in TmAu$_2$). With these parameters, we find that at the lowest temperatures, ferroquadrupolar order vanishes for strains of around $1\%$. }
\label{fig:mftj6} 
\end{figure}

In the previous section we described how group theoretic arguments guarantee the existence of three $E_g$ doublets when a $J=6$ system is embedded in a tetrgonal environment. The specific crystal field Hamiltonian took the form
\begin{align*}
\hat{H}_{CEF}=B^0_2\hat{O}^0_2+B^0_4\hat{O}^0_4+B^4_4\hat{O}^4_4+B^0_6\hat{O}^0_6+B^4_6\hat{O}^4_6
\end{align*}
where the $\hat{O}^m_l$ are the Stevens operators and $B^m_l$ are the corresponding coefficients, which in the case of TmAu$_2$ and TmAg$_2$ result in a ground state $E_g$ doublet. This crystal field Hamiltonian ignores quadrupole-Quadrupole interactions and quadrupole-lattice (magneto-elastic) interactions. In this section, we treat these terms within the framework of mean field theory. The corresponding mean field Hamiltonian takes the form
\begin{align}
\hat H_Q = 
- K_{B1g}\braket{\hat{O}^2_2}\hat{O}^2_2 + B_{B2g}\varepsilon_{xy}\hat{P}_{xy} 
\end{align}
Here, we have adopted the conventional notation $K_{B1g} = zJ_{ij}$ where $z$ is the number of neighbors for which $J_{ij}$ in Eq.~\ref{eq:fullh} is non-zero. The terms containing strains $\varepsilon_{x^2-y^2}$ and $\varepsilon_{xy}$ are the bilinear magneto-elastic terms, and the terms containing the quadrupole expectation values are the mean-field version of the quadrupole-quadrupole interaction terms, which arise due to a generalized version of the RKKY interaction\cite{Schmitt1985,MorinSchmitt88,SchmittQuadrupolar90, Morin1990Quadrupolar}.  $\hat{O}^2_2$ is the quadrupole operator, whose expectation value is the $B_{1g}$ order parameter, and is written in terms of angular momentum operators as $J_x^2-J_y^2$. Similarly, $P_{xy}$ is the quadrupole operator whose expectation value is the $B_{2g}$ order parameter, and is written as $\frac{1}{2}(J_xJ_y +J_yJ_x)$.  Assuming the material is free to relax, \textit{i.e.} the strain reaches the value that minimizes the free energy, the total Hamiltonian can be written as :
\begin{align}
\hat H_{total}(\braket{\hat{O}^2_2},\braket{P_{xy}}) = \hat H_{CEF} + \hat H_Q
\end{align}
The corresponding mean field equation for the $B_{1g}$ order parameter has the form
\begin{align}
\braket{\hat{O}^2_2}=  \frac{Tr[e^{-\beta \hat H_{total}(\braket{\hat{O}^2_2},\braket{P_{xy}})}\hat{O}^2_2] }{Tr[e^{-\beta \hat H_{total}(\braket{\hat{O}^2_2},\braket{P_{xy}})}]}
\end{align}
which we iterate numerically until self consistency is achieved. The results of this process for temperatures from 10K to 1K and transverse strains from 0 to 1.5\% is shown in the contour plot of the order parameter  $\braket{\hat{O}^2_2}$ vs. temperature and strain (Fig.~\ref{fig:mftj6}). Figure 3 of the main text shows the lowest 5 eigenenergies of the mean field Hamiltonian as a function of transverse strain $\varepsilon_{xy}$, for a fixed temperature of $T = 2$K. 

\subsection{Accuracy of transverse field Ising approximation for a $J=6$ system}

In Section~\ref{sec:effectiveH} we described how the low energy physics of the full $J=6$ system is that of a (pseudo) spin 1/2 transverse field Ising model. However, the presence of other CEF states in TmAg$_2$ and TmAu$_2$ means that this statement is approximate. This is because the ground state $E_{g}$ doublet will mix with the excited states in both the ordered phase and under large applied strains. An important question which we must therefore address is \textit{to what extent can we treat the low energy physics full $J=6$ system as that of the transverse field Ising model}? In this section, we show:
\begin{itemize}
\item The nature of the symmetry breaking (and hence the Ising character of the phase transition) is unaffected by higher energy states that admix with the ground state $E_g$ doublet.
\item The transverse field nature of $B_{2g}$ strain is still appropriate when we consider the full $J=6$ system. 
\item The low energy physics is effectively that of a two level system, (\textit{i.e.} an $S = 1/2$ pseudo-spin), as the character of the ground state wave function is predominantly that of the lowest energy CEF doublet ($\Gamma^{(1)}_5$).
\end{itemize}
Each of these points is discussed below.

\subsubsection{Ising universality in the $J=6$ system}
It is important to stress that regardless of whether we treat the essential degrees of freedom as the full $J=6$ system (which is already an approximation, since this describes just the lowest energy state of the spin orbit interaction), or as the projected pseudo-spin 1/2, the phase transition from tetragonal to quadrupolar order remains an Ising phase transition, and so all the critical phenomena are in the Ising universality class. The essential point is that the transition from a system with point group $D_{4h}$, to one with point group $D_{2h}$ is necessarily Ising in character. That is, a finite expectation value for the operator $\langle \hat{O}^2_2 \rangle$ implies that single $Z_2$ symmetry is being broken ($C_4$ rotations down to $C_2$ rotations).

\subsubsection{Transverse field nature of $B_{2g}$ strains}
While the projection to the ground state $E_g$ doublet makes manifest the relationship between operators $\hat{P}_{xy}$, $\hat{O}^2_2$, $\hat{H}_z$, and the Pauli matrices, we note that the essential physics of the transverse field Ising model is apparent even in the full $J=6$ description. In particular, the quadrupole operator $\hat{O}^2_2$ and the $B_{1g}$ operator $\hat{P}_{xy}$ do not commute. Including the operator magnetic field operator ($\hat{J}_z$) and using the definitions of Stevens operators in terms of angular momentum operators, we have the commutation relations:
\begin{equation}
\begin{aligned}
\left[\hat{J}_z,  \hat{O}^2_2 \right] &= 4 i \hat{P}_{xy}\\
\left[\hat{P}_{xy},  \hat{J}_z \right] &= i \hat{O}^2_2\\
\left[\hat{O}^2_2,  \hat{P}_{xy} \right] &= i \left( 2J(J+1) - 2\hat{J}^2_z - 1\right)\hat{J}_z
\end{aligned}
\end{equation}
The fact that $\hat{O}^2_2$ and $\hat{P}_{xy}$ do not commute means that these two operators do not share the same ground state. Increasing the $B_{2g}$ strain enhances quantum fluctuations between these differing ground states, and must eventually lead to a quantum phase transition, i.e. $B_{2g}$ strain acts in the same way as a transverse field in an Ising magnet. This is exactly the same argument that was presented in the main text for an orbital doublet  (pseudo-spin 1/2); here we have shown that it still applies even when the full $J=6$ manifold is considered. For completeness, we note that upon projecting these operator equations into the low energy doublet, we recover the typical spin $1/2$ commutation relations ($\mathfrak{su}(2)$ algebra).

\subsubsection{Validity of low energy limit}
\begin{figure}
\begin{center}
\includegraphics[width=0.5\textwidth]{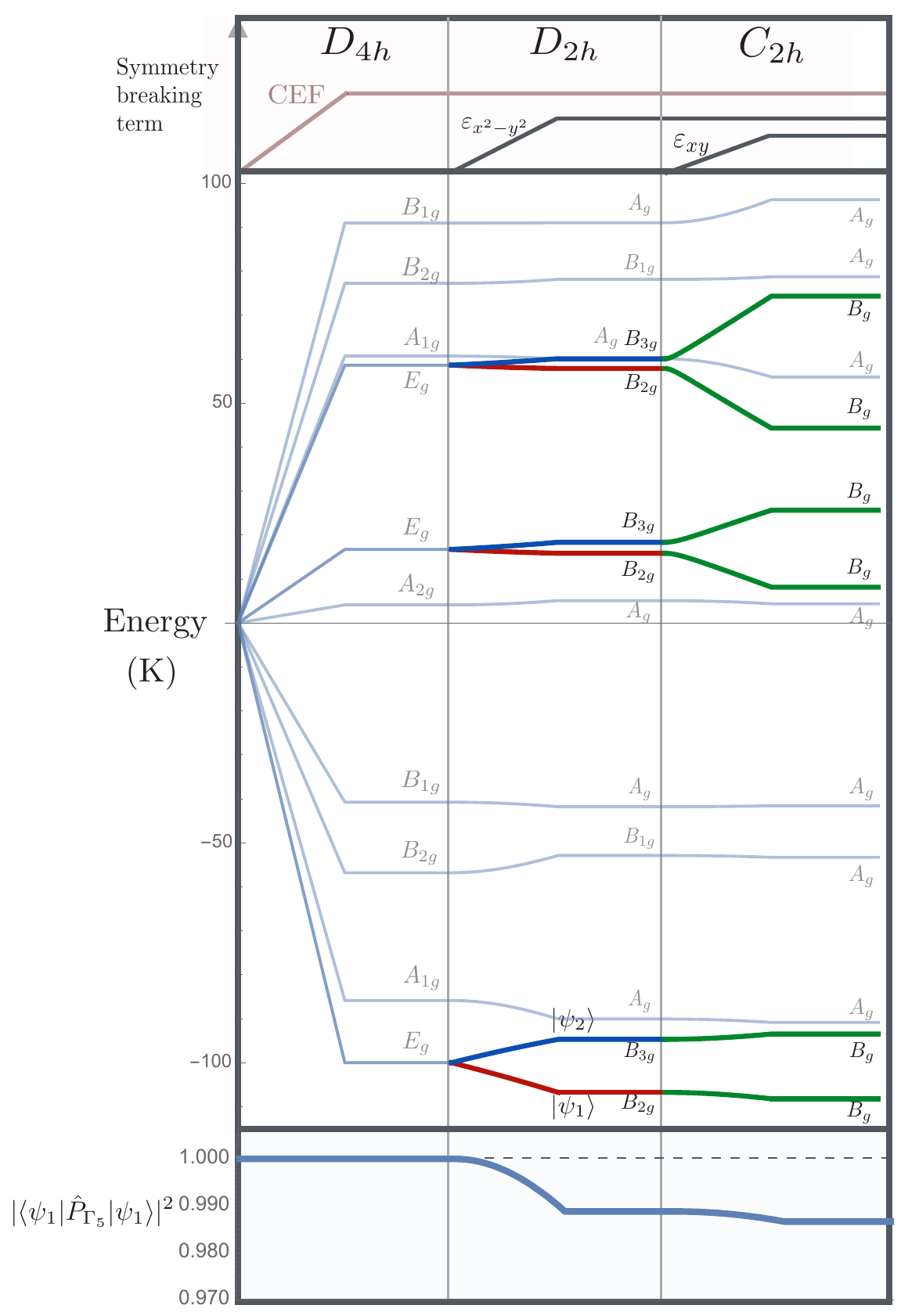} \label{fig:mixing}
 \caption{The evolution of the spectrum of a $J=6$ system (Eq.~\ref{eq:mfham}) under a series of symmetry breaking perturbations. The top panel shows the forms of onset for each symmetry breaking perturbation, and the resulting point group. The middle panel shows the evolution of energy eigenvalues, along with their symmetry labels in each regime. We have included labels for the representations to which each energy level belongs, and highlighted the evolution of the three $E_g$ doublets, $\Gamma^{(1)}_5, \Gamma^{(2)}_5, \Gamma^{(3)}_5$. The lowest panel shows the $\Gamma^{(1)}_5$ content of the lowest eigenstate $|\psi_1 \rangle$, i.e. the extent to which the ground state is comprised of only the lowest energy doublet of the CEF state. This overlap remains very close to 1 in the ferroquadrupolar phase, and only drops to 0.985 despite a relatively large $\varepsilon_{xy}$, suggesting that our approximation of pseudo-spin 1/2 physics is an excellent one.  }
 \end{center} 
\end{figure}

Our approximation of treating the system as a spin $1/2$ Ising model refers to projecting all operators and eigenstates into the basis of the lowest energy $E_g$ doublet (\textit{i.e.} the $\Gamma^{(1)}_5$ doublet) of the CEF Hamiltonian. This is an approximation due to the mixing of higher energy states with this lowest energy $\Gamma^{(1)}_5$ doublet under the perturbations of the quadrupole and transverse field operators $\hat{O}^2_2$ and $\hat{P}_{xy}$ respectively. However, one can show that $\hat{O}^2_2$ and $\hat{P}_{xy}$,  only mix states of $E_g$ character, which are well separated in energy from the ground state (see Fig. S5). Therefore, any deviation from spin $S = 1/2$ physics is suppressed by this energy gap.

The lower panel of Fig. S5 illustrates this point in a quantitative manner. We plot the overlap of the ground state with its projection in to the lowest energy $\Gamma^{(1)}_5$ doublet of the CEF Hamiltonian. In particular, denoting the eigenstates of low energy doublet of the CEF Hamiltonian as $|\psi^{\Gamma_5}_1\rangle$ and $|\psi^{\Gamma_5}_2 \rangle$, the projection operator into these states is
\begin{align}
\hat{P}_{\Gamma_5} = | \psi^{\Gamma_5}_1 \rangle \langle \psi^{\Gamma_5}_1 | +| \psi^{\Gamma_5}_2 \rangle \langle \psi^{\Gamma_5}_2 |  
\end{align}
We then consider the evolution of the ground state $|\psi_1\rangle$ of a mean field Hamiltonian in the full $J = 6$ basis,
\begin{align}
H_{MF} = H_{CEF} + \varepsilon_{x^2 - y^2} \hat{O}^2_2 + \varepsilon_{xy} \hat{P}_{xy}\label{eq:mfham}
\end{align}
Figure S5 (middle panel) shows the evolution of the spectrum of this Hamiltonian as various terms are turned on sequentially (upper panel). The maximum strength of each term is chosen to correspond to the maximum magnitudes of each operator in our self consistent mean field calculation (see Fig. S4). The lower panel shows the magnitude squared of the overlap 
\begin{align*}
\langle \psi_1 | \hat{P}_{\Gamma_5} | \psi_1 \rangle
\end{align*}
which is essentially the $\Gamma^{(1)}_5$ content of the ground state. As is clear from the figure, the `` spin $1/2$ " character of the ground state is very close to 1 throughout the evolution of the quadrupole and transverse field operators, indicating the validity of treating this system as a pseudo-spin $1/2$.

\end{document}